\documentstyle[prb,aps]{revtex}
\begin{document}
\draft
\tighten
\onecolumn
\title{Quantum interference from sums over closed paths 
for electrons \\
on a three-dimensional lattice in a magnetic field: \\
total energy, magnetic moment, and orbital susceptibility}
\author{Yeong-Lieh Lin\cite{lyl} and Franco Nori\cite{fnori}}
\address
{ Department of Physics, The University of Michigan, 
Ann Arbor, Michigan 48109-1120 }
\maketitle
\begin{abstract}
We study quantum interference effects due to electron motion on a
three-dimensional cubic lattice in a {\em continuously-tunable} magnetic 
field ${\bf B}$ of {\em arbitrary} orientation and magnitude. 
These effects arise from the interference
between magnetic phase factors associated with different electron
closed paths. The sums of these phase factors, called lattice 
path-integrals, are ``many-loop" 
generalizations of the standard ``one-loop" Aharonov-Bohm-type 
argument, where the electron wave function picks up
a phase factor $e^{i\Phi}$ each time it travels around a closed loop
enclosing a net flux $\Phi$. Our lattice path integral calculation 
enables us to obtain various important physical quantities through several 
different methods. The spirit of 
our approach follows Feynman's programme: to derive physical quantities 
in terms of ``sums over paths". From these lattice path-integrals we compute 
analytically, for several lengths of the electron path, 
the half-filled Fermi-sea ground-state energy, 
$E_T({\bf B})$, of noninteracting spinless electrons in a
cubic lattice. Our expressions for $E_T$ are valid for any strength of
the applied magnetic field in any direction. Moreover, we provide an
explicit derivation for the absolute minimum energy of the
flux state. For various field orientations, we also study the quantum
interference patterns and $E_T({\bf B})$ by exactly summing over 
$\sim 10^{29}$
closed paths in a cubic lattice, 
each one with its corresponding magnetic phase factor
representing the net flux enclosed by each path. Furthermore, an 
expression for the total kinetic energy $E_T({\bf B},\nu)$ for any 
electron filling $\nu$ close to one half is obtained. 
We also study in detail two experimentally important quantities: 
the magnetic moment $M({\bf B})$ and orbital susceptibility $\chi({\bf B})$ 
at half-filling, as well as the zero-field susceptibility 
$\chi(\mu)$ as a function of the Fermi energy $\mu$. 

\end{abstract}
\widetext

\section{Introduction}
The quantum behavior of noninteracting tight-binding electrons on a
two-dimensional (2D) lattice immersed in a perpendicular magnetic field
has attracted much attention due to its important role in diverse areas
of physics. For instance, the problem is intimately related to the
quantum Hall effect. During the past few years, the ground-state energy
of this system has also been intensively investigated\cite{hlrw,nori,lieb}
in connection with mean-field studies of the $t$-$J$ and Hubbard models
of high-$T_c$ superconductors. At half-filling, the absolute minimum of
the kinetic energy has a half flux quantum ($\Phi_{0}/2$) per 
plaquette---called the flux state. In addition, there are local minima at 
flux values equal to $\Phi_{0}/2m$ with $m > 1$. 

More recently, several groups have paid attention to the
three-dimensional (3D) case: the kinetic energy of a 3D noninteracting
electron gas under the influence of both a strong periodic potential
and a magnetic field. For instance, for several rational values of the
magnetic field, Skudlarski and Vignale\cite{vignale} analyzed the
changes in the ground-state properties induced by the addition of
hopping in the direction parallel to the uniform magnetic field.
Hasegawa\cite{hasegawa}  studied the density of states and the total
energy with the external field in the $(0,0,1)$, $(0,1,1)$, and $(1,1,1)$
directions for certain selected rational values of the resulting flux.
Kunszt and Zee\cite{zee} showed that for rational values of the flux 
the eigenvalue problem can be
reduced to a one-dimensional hopping in momentum space. They then
calculated the energy spectra and the density of states for various
selected rational values of the flux states. These three 
works\cite{vignale,hasegawa,zee} focused
on a {\em discrete} set of (rational) magnetic field values. Here we are 
interested in obtaining results that are {\em continuous} functions of the 
applied field---valid for both commensurate and incommensurate flux values.

In this work, we focus on the isotropic 3D cubic lattice, namely, the
hopping integrals in the $x$-, $y$-, and $z$-direction are the same and
taken to be equal to one. The system is described by the Hamiltonian
\begin{equation} 
H=\sum_{\langle ij \rangle}c_{i}^{\dag}c_{j}\exp(i
A_{ij}), 
\end{equation} 
where ${\langle ij \rangle}$ refers to nearest-neighbor sites, and 
the phase  $A_{ij}=2\pi\int_{i}^{j}{\bf A}{\cdot}d{\bf l}$
is $2\pi$ times the line integral of the vector potential along the bond 
from $i$ to
$j$. Throughout this paper, the flux quantum $\Phi_{0}=hc/e$ is set
equal to one.  The goal of this paper is to explore conceptually
different viewpoints and approaches to study this interesting problem.
Focusing on the hopping motion of electrons on the lattice, we first
study the quantum interference between the phase factors of electron
closed paths. This phenomenon is the source for the lowering of the
total energy in a magnetic field.

In Sec.~II, we examine the quantum interference effects originating
from lattice path-integrals---defined as sums over magnetic phase
factors on different electron closed paths. There, the physical meaning of
these lattice path-integrals ${\cal S}_{2l}$ and the techniques we use to 
compute them are
discussed in detail. Results for ${\cal S}_{2l}$ in an
arbitrarily oriented field of any strength are also presented. 
Furthermore, we obtain
the lattice path-integrals for several flux orientations by exactly
summing an enormous number ($\sim 10^{29}$) of closed paths, each one 
weighted by its corresponding acquired phase factor. 

In Sec.~III, we present an
analytical calculation of the total kinetic energy $E_T({\bf B})$ of the 
half-filled Fermi
sea of tight-binding electrons. We use two different approaches: one
based on a direct series expansion for the energy\cite{nori} and the
other on a moment expansion for the density of states.\cite{rice} These
two different methods yield the same results. Moreover, they provide different
insights into the problem. For an arbitrary field orientation, analytic 
results for the total energy, in terms of the lattice path-integrals, are 
obtained. It is found that the lowest energy state is reached when the 
fluxes per plaquette on each one of the $xy$-, $yz$- and $zx$-planes are 
equal to
$\Phi_{0}/2$. We also show the variations of the total energy $E_T$ as a 
function of the flux, for various orientations of the magnetic field.
Furthermore, we obtain an analytic expression for the total kinetic
energy $E_T({\bf B},\nu)$ for any filling $\nu$ close to one half. 

In Sec.~IV, we investigate
the magnetic moment $M({\bf B})$ and orbital susceptibility $\chi({\bf B})$ 
of this quantum system.
For various flux orientations, we also obtain
the zero-field susceptibility 
$\chi(\mu)$ as a function of the Fermi energy $\mu$. 
We show that the magnetic response, in the 
presence of a strong periodic potential, is significantly different from the 
familiar Landau diamagnetism. The zero-field susceptibility $\chi(\mu)$ as a 
function of the Fermi energy $\mu$ exhibits a large diamagnetic (i.e., 
negative $\chi$) response at very low electron filling. For increasing 
$\mu$, $\chi(\mu)$ increases and fluctuates around zero. Paramagnetism 
prevails for large $\mu$. Indeed, the orbital response is paramagnetic at 
and near half-filling ($\mu=0$). 
We obtain the field dependence of $M({\bf B})$---which 
can be regarded as a generalized current in a multiply-connected 
geometry---and $\chi({\bf B})$ at half-filling, and both show 
oscillations between para- and diamagnetic behaviors as a function of 
the flux. The frequencies of these oscillations, as functions of the flux, 
tend to decrease for increasing field.

In Sec.~V, we summarize our results.

\section{Quantum interference from sums over closed paths: lattice 
path-integrals}
\subsection{Physical interpretation}
The lattice path-integral of order $2l$ is defined as 
\begin{equation}
{\cal S}_{2l}\ \equiv\ \sum_{\stackrel{\scriptstyle\rm{All\ closed\ {\em
2l}-step}}{\scriptstyle\rm{lattice\ paths}\ \Gamma}}
\ e^{i\Phi_\Gamma}, 
\end{equation} 
where $\Phi_{\Gamma}$ is the sum
over phases of the bonds on the path $\Gamma$ of $2l$ steps starting
and ending at the same site.  Let $|\psi_i\rangle$ denote a localized
one-site electron state centered at site $i$. It is not difficult to
notice that ${\cal S}_{2l}$ corresponds precisely to the quantum
mechanical expectation value $\langle \psi_i |H^{2l}| \psi_i \rangle$,
which summarizes the contribution to the electron kinetic energy of
{\em all} closed paths of $2l$-steps.
 
The physical meaning of ${\cal S}_{2l}$ ($=\langle \psi_i |H^{2l}|
\psi_i \rangle$) thus becomes clear. The Hamiltonian $H$ is applied
$2l$ times to the initial state $|\psi_i\rangle$, resulting in the new
state $H^{2l}|\psi_i\rangle$ located at the end of the path traversing
$2l$ lattice bonds. Because of the presence of a magnetic field, a
magnetic phase factor $e^{i A_{ij}}$ is acquired by an electron when
hopping through two adjacent sites $i$ and $j$. ${\cal S}_{2l}$ is
nonzero only when the path ends at the starting site. In other words,
${\cal S}_{2l}$ is the sum of the contributions from all closed paths
of $2l$ steps starting and ending at the same site, each one weighted
by its corresponding phase factor $e^{i\Phi_\Gamma}$ where
$\Phi_\Gamma/2\pi$ is the {\em net\/} flux enclosed by the closed path
$\Gamma$.

It is important to stress that $\Phi_{\Gamma}$ depends crucially on the
traveling route of the path. For instance, $\Phi_{\Gamma}$ will be
positive (negative) by traversing a polygon loop counterclockwise
(clockwise). Therefore, quantum interference information contained in ${\cal
S}_{2l}$ arises because the phase factors of different closed paths,
including those from all kinds of distinct loops and separate
contributions from the same loop, interfere with each other. Sometimes,
the phases corresponding to subloops of a main path cancel.

\subsection{Analytical computational formalism}
We will now compute the lattice path-integrals ${\cal S}_{2l}$. This is
a difficult task since ${\cal S}_{2l}$ involves an enormous number of
different paths (growing rapidly when the order increases), each one
given by its corresponding net magnetic phase factor. We have
considerably simplified this calculation by successively iterating the
recursion relation and analyzing the symmetries of the problem. 

We consider a unit spacing for the cubic lattice.  The vector potential of
a general magnetic field ${\bf B}=(B_{x},B_{y},B_{z})$ can be written
as 
$${\bf A}=\frac{1}{2}(zB_y-yB_z, xB_z-zB_x, yB_x-xB_y).$$ 
Let $a/2\pi$,
$b/2\pi$, and $c/2\pi$ represent the three fluxes through the respective
elementary plaquettes on the $yz$-, $zx$- and $xy$-planes. Thus, a
flux configuration is specified by $(a,b,c)$. 

From the definition of ${\cal S}_{2l}$, it is clear that: (i) ${\cal
S}_{0}$ obviously equals $1$, (ii)  ${\cal S}_{2l+1}$ are always zero
because there is no path with an odd number of steps for returning an
electron to its initial site, (iii) the ${\cal S}_{2l}$'s are gauge
invariant, and (iv) in the absence of the magnetic field, ${\cal
S}_{2l}$ are just the total number of $2l$-step paths on the cubic
lattice starting and ending at the same site.

First, it is instructive to evaluate the first two lattice path-integrals. 
This will help clarify their physical meaning. 
For the two-step closed paths, and starting from any initial site, 
the electron retraces its first step on one 
of the six bonds connecting the initial site with its adjacent sites. 
This process can be designated symbolically by 
$(\mbox{\boldmath $\cdot$}\mbox{\boldmath $\leftrightarrow$})$, 
where the dot ($\mbox{\boldmath $\cdot$}$) indicates the initial site. 
The flux enclosed, of course, is zero. From these six closed paths of 
two steps each, we obtain
$${\cal S}_2=6\,(\mbox{\boldmath $\cdot$}\mbox{\boldmath $\leftrightarrow$})
=6\,e^{i0}=6=z,$$
where $z$ stands for the coordination number of the cubic lattice. 

For the four-step closed paths, we need to consider four different 
possibilities: (1) the electron retraces twice 
each one of the six bonds connecting the initial site with its 
nearest-neighboring sites 
$(\mbox{\boldmath $\cdot$}\mbox{\boldmath 
$\stackrel{{\textstyle \leftrightarrow}}{\leftrightarrow}$})$; 
(2) after retracing once  one 
of the six bonds connecting to the initial site, the electron 
retraces two steps on one of the other five bonds 
$(\mbox{\boldmath $\leftrightarrow$}\mbox{\boldmath $\cdot$}\mbox{\boldmath 
$\leftrightarrow$})$; (3) hopping first to one of the six adjacent sites, 
the electron retraces once one of the other five bonds (the one connecting 
to the initial site is excluded) and then returns to the original site 
$(\mbox{\boldmath $\cdot$}\mbox{\boldmath 
$\stackrel{{\textstyle \leftarrow}}{\rightarrow}$}\mbox{\boldmath 
$\leftrightarrow$})$; and (4) the electron traverses either 
counterclockwise or clockwise on one of the four elementary square cells 
(connecting to the initial site) 
on the $yz$-, $zx$- and $xy$-planes respectively
[$(\mbox{\boldmath $\cdot$}\mbox{\boldmath $\stackrel{\leftarrow}{\Box}$}+
\mbox{\boldmath $\cdot$}\mbox{\boldmath $\stackrel{\rightarrow}{\Box}$})_{yz}+
(\mbox{\boldmath $\cdot$}\mbox{\boldmath $\stackrel{\leftarrow}{\Box}$}+
\mbox{\boldmath $\cdot$}\mbox{\boldmath $\stackrel{\rightarrow}{\Box}$})_{zx}+
(\mbox{\boldmath $\cdot$}\mbox{\boldmath $\stackrel{\leftarrow}{\Box}$}+
\mbox{\boldmath$\cdot$}\mbox{\boldmath$\stackrel{\rightarrow}{\Box}$})_
{xy}]$. Thus
\begin{eqnarray*}
{\cal S}_4&=&6\,(\mbox{\boldmath $\cdot$}\mbox{\boldmath 
$\stackrel{{\textstyle \leftrightarrow}}{\leftrightarrow}$})+
30\,(\mbox{\boldmath $\leftrightarrow$}\mbox{\boldmath $\cdot$}
\mbox{\boldmath $\leftrightarrow$})+
30\,(\mbox{\boldmath $\cdot$}\mbox{\boldmath $\stackrel{{\textstyle 
\leftarrow}}{\rightarrow}$}\mbox{\boldmath $\leftrightarrow$}) \\
& &\mbox{}+4\,\left[(\mbox{\boldmath $\cdot$}\mbox{\boldmath 
$\stackrel{\leftarrow}{\Box}$}+
\mbox{\boldmath $\cdot$}\mbox{\boldmath $\stackrel{\rightarrow}{\Box}$})_{yz}+
(\mbox{\boldmath $\cdot$}\mbox{\boldmath $\stackrel{\leftarrow}{\Box}$}+
\mbox{\boldmath $\cdot$}\mbox{\boldmath $\stackrel{\rightarrow}{\Box}$})_{zx}+
(\mbox{\boldmath $\cdot$}\mbox{\boldmath $\stackrel{\leftarrow}{\Box}$}+
\mbox{\boldmath$\cdot$}
\mbox{\boldmath$\stackrel{\rightarrow}{\Box}$})_{xy}\right] \\
&=&(6+30+30)\,e^{i0} \\
& &\mbox{}+4\,[(e^{ia}+e^{-ia})+(e^{ib}+e^{-ib})
+(e^{ic}+e^{-ic})] \\
&=&66+8\,(\cos a+\cos b+\cos c).
\end{eqnarray*}

To compute ${\cal S}_{2l}$ in a systematic manner, we first define the 
quantity $S^{(t)}_{p,q,r}$, which is the sum over all possible paths of 
$t$ steps on which an electron may hop from the origin $(0,0,0)$ to site 
$(p,q,r)$. From the definition of 
$S^{(t)}_{p,q,r}$, it is straightforward to construct the following 
recurrence relation for $S^{(t+1)}_{p,q,r}$
\begin{equation}
S^{(t+1)}_{p,q,r}=\exp\!\left(\pm i\,\frac{qc-rb}{2}\right)
S^{(t)}_{p\pm 1,q,r}+\exp\!\left(\pm i\,\frac{ra-pc}{2}\right)
S^{(t)}_{p,q\pm 1,r}
+\exp\!\left(\pm i\,\frac{pb-qa}{2}\right)S^{(t)}_{p,q,r\pm 1}.
\end{equation} 
This equation states that the site $(p,q,r)$ can be reached by taking the 
$(t+1)$th step from the six nearest-neighboring sites. The factors in 
front of the $S^{(t)}$'s account for the presence of the magnetic field. 
By recursively using Eq.~(3), we obtain the recurrence relation for
$S^{(t+2)}_{p,q,r}$ as 
\begin{eqnarray}
S^{(t+2)}_{p,q,r}&=&6\,S^{(t)}_{p,q,r}
+\exp[\,\pm i\,(qc-rb)\,]\,S^{(t)}_{p\pm 2,q,r}
+\exp[\,\pm i\,(ra-pc)\,]\,S^{(t)}_{p,q\pm 2,r}
+\exp[\,\pm i\,(pb-qa)\,]\,S^{(t)}_{p,q,r\pm 2} \nonumber \\
& &\mbox{}+2\cos\left(\frac{c}{2}\right)\left[
\exp\!\left(\pm i\,\frac{r(a-b)-(p-q)c}{2}\right)S^{(t)}_{p\pm 1,q\pm 1,r}+
\exp\!\left(\mp i\,\frac{r(a+b)-(p+q)c}{2}\right)S^{(t)}_{p\pm 1,q\mp 1,r}
\right]   \nonumber \\
& &\mbox{}+2\cos\left(\frac{b}{2}\right)\left[
\exp\!\left(\pm i\,\frac{q(c-a)-(r-p)b}{2}\right)S^{(t)}_{p\pm 1,q,r \pm 1}+
\exp\!\left(\mp i\,\frac{q(c+a)-(r+p)b}{2}\right)S^{(t)}_{p\pm 1,q,r \mp 1}
\right]   \nonumber \\
& &\mbox{}+2\cos\left(\frac{a}{2}\right)\left[
\exp\!\left(\pm i\,\frac{p(b-c)-(q-r)a}{2}\right)S^{(t)}_{p,q\pm 1,r \pm 1}+
\exp\!\left(\mp i\,\frac{p(b+c)-(q+r)a}{2}\right)S^{(t)}_{p,q\pm 1,r \mp 1}
\right].
\end{eqnarray}

We will now use Eq.~(4) as the basic recurrence relation in our iteration 
scheme. Without loss of generality and for convenience, we choose the 
origin $(0,0,0)$ to be our starting site. The initial conditions then read 
$S^{(0)}_{0,0,0}=1$ and $S^{(0)}_{p,q,r}=0$ for other $(p,q,r)$'s. 
It is evident that ${\cal S}_{2l}$ is just equal to $S^{(2l)}_{0,0,0}$. 
Putting $t=0$ in Eq.~(4), we directly have  
$S^{(2)}_{0,0,0}=6$,
\begin{eqnarray*} 
S^{(2)}_{1,1,0}&=&S^{(2)}_{1,-1,0}=S^{(2)}_{-1,1,0}
=S^{(2)}_{-1,-1,0}=2\,\cos\left(\frac{c}{2}\right), \\ 
S^{(2)}_{1,0,1}&=&S^{(2)}_{1,0,-1}
=S^{(2)}_{-1,0,1}=S^{(2)}_{-1,0,-1}=2\,\cos\left(\frac{b}{2}\right), \\ 
S^{(2)}_{0,1,1}&=&S^{(2)}_{0,1,-1}
=S^{(2)}_{0,-1,1}=S^{(2)}_{0,-1,-1}=2\,\cos\left(\frac{a}{2}\right),
\end{eqnarray*}
and 
$$S^{(2)}_{2,0,0}=S^{(2)}_{-2,0,0}
=S^{(2)}_{0,2,0}=S^{(2)}_{0,-2,0}
=S^{(2)}_{0,0,2}=S^{(2)}_{0,0,-2}=1.$$ Using the above results for the 
$S^{(2)}_{p,q,r}$'s and putting $t=2$ in Eq.~(4), we then obtain the 
$S^{(4)}_{p,q,r}$'s. In general, by following this procedure we can obtain 
$S^{(2l)}_{0,0,0}$ to any desired $l$. The properties discussed below 
will make the computation of $S^{(2l)}_{0,0,0}$ quite efficient.

For a given $l$, a nonzero $S^{(2l)}_{p,q,r}$ exists only on those $p$, $q$, 
and $r$ satisfying $|p|+|q|+|r|=0,2,\ldots,2l$. This stems from the fact 
that only the sites $(p,q,r)$ satisfying this condition can be reached by 
an electron after $2l$ hops from $(0,0,0)$. The total number of these sites 
is $$N^{(2l)}_{T}=\frac{16l^3+24l^2+14l+3}{3}.$$ 
Among the $S^{(2l)}_{p,q,r}$'s, we find 
the following symmetries hold for any $l$. We omit 
the superscript $(2l)$ below.
\begin{eqnarray*} 
S_{p,q,r}(a,b,c)&=&S_{-p,-q,-r}(a,b,c), \\
S_{-p,q,r}(a,b,c)&=&S_{p,-q,-r}(a,b,c), \\
S_{p,-q,r}(a,b,c)&=&S_{-p,q,-r}(a,b,c), \\
S_{p,q,-r}(a,b,c)&=&S_{-p,-q,r}(a,b,c),  
\end{eqnarray*}
$$S_{p,q,r}(a,b,c)=S_{-p,q,r}(-a,b,c)=S_{p,-q,r}(a,-b,c)=S_{p,q,-r}(a,b,-c),$$
and
$$S_{p,q,r}(a,b,c)=S_{p,r,q}(a,c,b)=S_{r,p,q}(b,c,a)=S_{q,p,r}(b,a,c)=
S_{q,r,p}(c,a,b)=S_{r,q,p}(c,b,a).$$
We can therefore reduce our calculation to the $S^{(2l)}$'s at sites 
$(p,q,r)$ in the first octant with $p\geq q\geq r\geq 0$. Thus, the 
total number of the independent $S^{(2l)}_{p,q,r}$ is 
$$N_{I}^{(2l)}=\frac{l^3+6l^2+12l+9-n}{9}$$ 
for $l=3m+n$, where $m$ are non-negative integers and $n=0,1,2.$ 
Thus the number of the 
$S^{(2l)}$'s to be computed is significantly reduced 
(about a factor of $48$) compared to that of the whole $S^{(2l)}_{p,q,r}$.

Finally, it is worthwhile to note that to obtain the lattice path-integrals 
up to ${\cal S}_{2L}$, it is sufficient to compute the 
$S^{(L)}_{p,q,r}$'s for $0\leq p+q+r \leq L$ when $L$ is even and the 
$S^{(L+1)}_{p,q,r}$'s for $0\leq p+q+r \leq L-1$ when $L$ is odd. 
To be more specific, 
in doing the iteration of $S^{(2l)}_{p,q,r}$ for $L/2 \leq l \leq L$, 
we need only to concern ourselves with those $S^{(2l)}_{p,q,r}$'s with 
$p+q+r=0,2,\ldots,2(L-l)$.

\subsection{Results for lattice path-integrals with general flux orientations}
For the general flux orientation $(a,b,c)$, we have obtained 
${\cal S}_{2l}$ up to $2l=20$. Here: $\sum_{(\alpha)}$ denote sums over 
$\alpha=a, b, c$; \,$\sum_{(\alpha\,\beta)}$ denote sums over 
$(\alpha\,\beta)=(a\,b), (b\,c), (c\,a)$; and \,
$\sum_{(\alpha\,\beta\,\gamma)}$ denote sums over 
$(\alpha\,\beta\,\gamma)=(a\,b\,c), (b\,c\,a), (c\,a\,b)$. 
Also, for instance, the term $\cos(\alpha \pm \beta)$ means 
$\cos(\alpha + \beta)+\cos(\alpha - \beta)$. Below we present the results for 
${\cal S}_{4}$, ${\cal S}_{6}$, ${\cal S}_{8}$, and ${\cal S}_{10}$.
\begin{eqnarray*}
{\cal S}_{4} &=& 66+8\sum_{(\alpha)}\cos\alpha, \\
{\cal S}_{6} &=& 876+\sum_{(\alpha)}[\,240\cos\alpha+24\cos2\alpha\,]
+24\sum_{(\alpha\,\beta)}\cos(\alpha\pm \beta)+12\cos(a+b+c)
+12\sum_{(\alpha\,\beta\,\gamma)}\cos(\alpha+\beta-\gamma), \\
{\cal S}_{8} &=& 12978+\sum_{(\alpha)}\left[\,5632\cos\alpha+1000\cos2\alpha
+96\cos3\alpha+16\cos4\alpha\,\right] \\ 
& &\mbox{}+\sum_{(\alpha\,\beta)}[\,1120\cos(\alpha\pm \beta)
+112\cos(2\alpha\pm \beta)+112\cos(\alpha\pm 2\beta)
+32\cos(2\alpha\pm 2\beta)\,]+576\cos(a+b+c)   \\
& &\mbox{}+\sum_{(\alpha\,\beta\,\gamma)}
[\,576\cos(\alpha+\beta-\gamma)+64\cos(2\alpha +\beta\pm \gamma)
+64\cos(2\alpha -\beta\pm \gamma) \\
& &\mbox{}+16\cos(\alpha+2\beta\pm 2\gamma)
+16\cos(\alpha-2\beta\pm 2\gamma)\,], \\
{\cal S}_{10} &=& 208836+\sum_{(\alpha)}\left[\,124080\cos\alpha
+30040\cos2\alpha
+5040\cos3\alpha+1160\cos4\alpha+160\cos5\alpha+40\cos6\alpha\,\right] \\ 
& &\mbox{}+\sum_{(\alpha\,\beta)}[\,36680\cos(\alpha\pm \beta)
+6800\cos(2\alpha\pm \beta)+6800\cos(\alpha\pm 2\beta)
+2160\cos(2\alpha\pm 2\beta) \\
& &\mbox{}+600\cos(3\alpha\pm \beta)+600\cos(\alpha\pm 3\beta)
+240\cos(3\alpha\pm 2\beta)+240\cos(2\alpha\pm 3\beta)
+80\cos(4\alpha\pm \beta) \\
& &\mbox{}+80\cos(\alpha\pm 4\beta)+60\cos(2\alpha\pm 4\beta)
+60\cos(2\alpha\pm 4\beta)+40\cos(3\alpha\pm 3\beta)\,]+19860\cos(a+b+c) \\
& &\mbox{}+\sum_{(\alpha\,\beta\,\gamma)}
[\,19860\cos(\alpha+\beta-\gamma)+4040\cos(2\alpha +\beta\pm \gamma)
+4040\cos(2\alpha -\beta\pm \gamma) \\
& &\mbox{}+1280\cos(\alpha+2\beta\pm 2\gamma)
+1280\cos(\alpha-2\beta\pm 2\gamma)+400\cos(3\alpha+\beta\pm \gamma)
+400\cos(3\alpha-\beta\pm \gamma) \\
& &\mbox{}+160\cos(\alpha+2\beta\pm 3\gamma)
+160\cos(\alpha-2\beta\pm 3\gamma)+160\cos(\alpha+3\beta\pm 2\gamma)
+160\cos(\alpha-3\beta\pm 2\gamma) \\
& &\mbox{}+40\cos(4\alpha+\beta\pm \gamma)
+40\cos(4\alpha-\beta\pm \gamma)+40\cos(3\alpha+2\beta\pm 2\gamma)
+40\cos(3\alpha-2\beta\pm 2\gamma) \\
& &\mbox{}+40\cos(\alpha+4\beta\pm 2\gamma)
+40\cos(\alpha-4\beta\pm 2\gamma)+40\cos(\alpha+2\beta\pm 4\gamma)
+40\cos(\alpha-2\beta\pm 4\gamma) \\
& &\mbox{}+20\cos(4\alpha+2\beta\pm 2\gamma)
+20\cos(4\alpha-2\beta\pm 2\gamma)+20\cos(\alpha+3\beta\pm 3\gamma)
+20\cos(\alpha-3\beta\pm 3\gamma)\,].
\end{eqnarray*}
We have also worked out a considerable number of lattice path-integrals 
(up to the $40$th order) for various special orientations of the flux. 
The results for ${\cal S}_{4}$, ${\cal S}_{6}$,$\ldots$, ${\cal S}_{12}$ 
are listed in Table I. The results for ${\cal S}_{14}$, 
${\cal S}_{16}$,$\ldots$, ${\cal S}_{40}$ are not presented.

\section{Total kinetic energy}
We now proceed to calculate the total kinetic energy $E_T$ of the half-filled 
Fermi sea. We do this by using a series expansion of the total energy. 
We also present an alternative approach based on a moment expansion of 
the density of states. Through these two different approaches, we show 
that our goal here---the calculation of $E_T({\bf B})$---is 
reduced to the evaluation of the lattice path-integrals ${\cal S}_{2l}$.

\subsection{Series expansion for the total energy}
Let us work on the $\{|\psi_i\rangle \}$ basis. At half-filling, 
the total kinetic energy of noninteracting spinless electrons is the sum of 
the lowest $N/2$ eigenvalues, where $N$ is the total number of 
sites. Since the energy spectrum of the cubic lattice is symmetric under 
$\{E\} \rightarrow \{-E\}$, we can write the total energy per site as 
\begin{equation}
E_{T}=\frac{1}{N}\sum_{E<0} E
=-\frac{z}{2N}\ {\rm Tr}\ \left|\frac{H_{0}}{z}\right|.
\end{equation}
Here \,Tr\, denotes the trace and $H_{0}$ is the corresponding diagonalized 
Hamiltonian of $H$. Notice that the absolute value is typically defined 
for scalar numbers. In this case, $|H_{0}/z|$ refers to an operator 
obtained by taking the absolute value of every matrix element of the 
operator $H_{0}/z$. Noting that 
$-I \leq |H_0/z| \leq I$, we expand $|H_{0}/z|$ into a series in terms of 
Chebyshev polynomials $T_{2k}(H_0/z)$ as
\begin{equation}
\left|\frac{H_0}{z}\right|=\frac{2}{\pi}\,I+\frac{4}{\pi}\sum_{n=1}^{\infty}
\frac{(-1)^{n+1}}{4n^{2}-1}\,T_{2n}\left(\frac{H_0}{z}\right),
\end{equation}
where $I$ represents the identity operator. Exploiting the equality
\begin{equation}
T_{2n}\left(\frac{H_0}{z}\right)=(-1)^{n}\,n\,
\sum_{l=0}^{n}\,\Omega_{l}\,H_{0}^{2l},
\end{equation}
where
$$\Omega_{l}=\left(\frac{-4}{z^2}\right)^l\frac{(n+l-1)!}{(2l)!\,(n-l)!},$$
we obtain from Eqs.~(5-7)
\begin{equation}
E_{T}=-\frac{z}{\pi}\left[1-\frac{2}{N}\sum_{n=1}^{\infty}\frac{n}{4n^{2}-1}
\left(\sum_{l=0}^{n}\Omega_{l}\,({\rm Tr}\,H^{2l})\right)\right],
\end{equation}
where we have replaced ${\rm Tr}(H_{0}^{2l})$ by 
${\rm Tr}(H^{2l})$, as they are equal to each other.

Assuming periodic boundary conditions on the lattices, we have
\begin{equation}
{\rm Tr}\,H^{2l}= N\,\langle\psi_i|H^{2l}|\psi_i\rangle.
\end{equation}
The total kinetic energy per site is then given by
\begin{equation}
E_{T}(a,b,c)=-\frac{z}{\pi}\left[1-2\sum_{n=1}^{L}\frac{n}{4n^{2}-1}
\left(\sum_{l=0}^{n}\Omega_{l}\,{\cal S}_{2l}(a,b,c)\right)\right],
\end{equation}
where we have assumed the highest order of the lattice path-integral 
obtained is $2L$. The above result is exact when $L=\infty$. Truncations at 
high values of $L$ provide excellent approximations to the $L=\infty$ case, 
because the terms corresponding to $L$ larger than either $4$, $5$, or 
$6$ (depending on the orientation of the field) are negligible. Our results 
for the general field orientation $(a,b,c)$ of $E_T$ include all terms up to 
$L=10$. For the
special flux configurations $(0,0,\phi)$, $(0,\phi,\phi)$, 
$(\phi,\phi,\phi)$, $(\phi,\phi,-2\phi)$, and $(\pi,\pi,\phi)$, 
we obtain all terms up to $L=20$. 
Terms up to $L=10$ (corresponding to ${\cal S}_{20}$) include contributions 
originating from $\sim 10^{14}$ three-dimensional closed paths, while terms 
up to $L=20$ (corresponding to ${\cal S}_{40}$) 
include contributions coming from $\sim 10^{29}$ closed paths 
in a 3D cubic lattice. Each one of these closed paths is weighted 
by its corresponding magnetic phase factor.

Note that, by construction, the lattice path-integrals are local quantities 
and are valid for any value of $\phi$. The periodic boundary conditions 
make the lattice path-integrals homogeneous (i.e., translationally invariant) 
and are unrelated to the imposition of a Bloch theorem. In contrast to most 
works studying tight-binding electrons in a magnetic field, here we never 
invoke a $k-$space or reciprocal space: our sums are all defined in direct 
space.

\subsection{Moment expansion for the density of states}
Here we analytically apply the method of moment expansion to obtain the 
total energy $E_T$. The starting point now is the one-particle Green's 
function 
$G_{ii}(E)\equiv\langle \psi_i|(E-H)^{-1}|\psi_i\rangle$, which can be 
expressed as
\begin{equation}
G_{ii}(E)=\frac{1}{E}\,\sum_{l=0}^{\infty}\,
\frac{\langle \psi_i|H^{2l}|\psi_i\rangle}
{E^{2l}}=\frac{1}{E}\,\sum_{l=0}^{\infty}\,\frac{{\cal S}_{2l}}
{E^{2l}}.
\end{equation}
The nonzero $2l$-th moment ${\cal M}_{2l}$ of the density of states 
$\rho(E)$ is given by
\begin{equation}
{\cal M}_{2l}=\int_{-z}^{z}\left(\frac{E}{z}\right)^{2l}\rho(E)\,dE.
\end{equation}
Taking into account the relation between 
$\rho(E)$ and the imaginary part of the Green's function, it can be derived 
that
\begin{equation}
{\cal M}_{2l}=\frac{{\cal S}_{2l}}{z^{2l}}.
\end{equation}
After the change of variables $\omega=E/z$, Eq.~(12) can be rewritten as
\begin{equation}
{\cal M}_{2l}=\int_{-1}^{1}\omega^{2l}\,\tilde{\rho}(\omega)\,d\omega,
\end{equation}
where $\tilde{\rho}(\omega)=z\,\rho(\omega z)$. 

We now expand $\tilde{\rho}(\omega)$ in a series of Chebyshev polynomials 
weighted by $1/\sqrt{1-\omega^2}$ as
\begin{equation}
\tilde{\rho}(\omega)=\sum_{k=0}^{\infty}C_{2k}\,\frac{T_{2k}(\omega)}
{\sqrt{1-\omega^2}}.
\end{equation}
Substituting Eq.~(15) into Eq.~(14) and using for $l \geq k$
\begin{equation}
\int_{-1}^{1} \frac{\omega^{2l}\,T_{2k}(\omega)}{\sqrt{1-\omega^2}}\,d\omega 
\,\equiv\, I_{2l,2k}\,=\,\frac{(2l)!\,\pi}{4^l\,(l-k)!\,(l+k)!},
\end{equation}
we obtain
\begin{equation}
{\cal M}_{2l}=\sum_{k=0}^{l}C_{2k}\,I_{2l,2k}\,.
\end{equation}
Given the moments up to ${\cal M}_{2L}$, the coefficients $C_{2k}$, for 
$k=0,\ldots,L$, can then be 
exactly determined one by one in terms of the moments.
For instance, 
$$C_0=\frac{{\cal M}_{0}}{I_{0,0}}=\frac{1}{\pi},$$
$$C_2=\frac{{\cal M}_{2}-C_0\,I_{2,0}}{I_{2,2}}=-\frac{4}{3\pi},$$ 
and
\begin{eqnarray*} 
C_4&=&\frac{{\cal M}_{4}-C_0\,I_{4,0}-C_2\,I_{4,2}}{I_{4,4}} \\
&=&\frac{4}{27\pi}\left[1+\frac{2}{3}(\cos a+\cos b+\cos c)\right].
\end{eqnarray*}
In general, every $C_{2k}$ can be computed from the previous ones by
\begin{equation}
C_{2k}=\frac{1}{I_{2k,2k}}\left(
{\cal M}_{2k}-\sum_{i=0}^{k-1}C_{2i}\,I_{2k,2i}\right).
\end{equation}
Noting that at half-filling the total energy  $E_{T}$ is
\begin{equation}
E_{T}=\int_{-z}^{0}E\,\rho(E)\,dE,
\end{equation}
and using the result
$$\int_{-1}^{0} \frac{\omega\,T_{2k}(\omega)}{\sqrt{1-\omega^2}}\,d\omega
=\frac{(-1)^k}{(2k-1)\,(2k+1)},$$
we therefore obtain the following expression for the total kinetic energy
\begin{equation}
E_{T}(a,b,c)=z\,\sum_{k=0}^{L}\frac{(-1)^k}{(2k-1)\,(2k+1)}\,C_{2k}(a,b,c).
\end{equation}
It is interesting to notice that Eqs.~(10) and (20) lead to the same result 
for the total energy $E_T$, even though we used quite different approaches. 
Both results are exact when $L=\infty$. However, terms with $k$ larger 
than $4$, $5$, or $6$ (depending on the field orientation) 
provide negligible contributions to the sum. Our results for the general 
field orientation $(a,b,c)$ of $E_T(a,b,c)$ include all terms up to $L=10$, 
while for the five special flux configurations studied below (Sec.~III.C), 
we include all terms up to $L=20$.

\subsection{The lowest-energy flux state and $E_T$ for various flux 
orientations}
Substituting the expressions for the ${\cal S}_{2l}(a,b,c)$'s into Eq.~(10) 
or the results for the $C_{2l}(a,b,c)$'s into Eq.~(20), 
we find that the lowest energy state 
is reached when $a=b=c=\pi$ (namely, a half flux quantum per plaquette on 
the $yz$-, $zx$- and $xy$-planes). We thus provide an {\em explicit 
analytic derivation for the absolute minimum energy of the flux state}. 
The result is consistent with a theorem proved by Lieb.\cite{lieb} 

We also obtain the total energy $E_T(\phi)$ as a {\em continuous} function 
of the flux $\phi$ for the following flux orientations: $(0,0,\phi)$, 
$(0,\phi,\phi)$, $(\phi,\phi,\phi)$, $(\phi,\phi,-2\phi)$, and the 
asymmetric case $(\pi,\pi,\phi)$. Notice that the field direction of 
$(\phi,\phi,\phi)$ is perpendicular to that of $(\phi,\phi,-2\phi)$. 
These results are plotted in Fig.~1. They are calculated through Eq.~(20) 
by using their respective $C_{2k}$'s for $2k=0, 2, 4,\ldots, 40.$ Thus, 
we have added the contributions 
of $\sim 10^{29}$ electron closed paths in a 3D cubic lattice, each one 
weighted by its magnetic phase factor.

For comparison purposes, we also present $E_{T}^{(2D)}(\phi)$ for the 
2D case, obtained by using the corresponding $C^{(2D)}_{2k}$'s up to 
$C^{(2D)}_{76}$.  This corresponds to summing over $\sim 10^{44}$ electron 
closed paths in a 2D square lattice, each one with its corresponding 
phase factor. Note that the absolute minimum of $E_T(\phi)$ occurs at 
$\phi=\pi$ in all of these flux orientations, except for the orientation 
$(0,0,\phi)$ in 3D. It becomes clear that hopping in an extra dimension 
drastically changes the properties found in strictly two-dimensional systems.

In Fig.~2, we plot the total kinetic energy of electrons at half-filling 
$E_T$ for the 3D cubic lattice for various field orientations versus the 
order of the highest-order lattice path-integral used to calculate them. 
For comparison, we also show the numerical values of $E_T$ obtained 
in Ref.~5, which mostly has only three significant figures. The $E_T$ reach 
steady values which are consistent, within $\sim 2\%$, with numerical ones 
for lattice path-integrals of order $2L$ equal to either $4$, $6$, or $8$, 
depending on the orientation of the field. Higher order lattice 
path-integrals provide negligible contributions to $E_T$. 
For instance, for $(0,0,0)$ both approaches give the same result, within 
$\sim 0.5\%$, for $2L\geq 4$.

\subsection{Total energy for any filling}
The approach described in section III.B can be directly generalized to 
calculate the total energy $E_{T}(\nu)$ for {\em any} electron filling 
$\nu$ $(0 \leq \nu \leq 1/2)$ as
\begin{equation}
E_{T}(\nu)=\int_{-z}^{\mu}E\,\rho(E)\,dE
=z\int_{-1}^{\mu/z}\omega\,\tilde{\rho}(\omega)
\,d\omega,
\end{equation}
where the Fermi energy $\mu$ is determined by
\begin{equation}
\nu=\int_{-z}^{\mu}\rho(E)\,dE
=\int_{-1}^{\mu/z}\tilde{\rho}(\omega)\,d\omega.
\end{equation}
Utilizing Eq.~(15) and carrying out the integrals, we obtain
\begin{equation}
E_{T}(\nu)=-\frac{z}{\pi}\sin\theta-\frac{z}{2}\sum_{k=1}^{L}
C_{2k}\left[\frac{\sin[(2k-1)\theta]}{2k-1}
+\frac{\sin[(2k+1)\theta]}{2k+1}\right],
\end{equation}
where $\theta=\arccos(\mu/z)$ $(\pi \geq \theta \geq \pi/2)$, and given 
$\nu$, $\theta$ can be solved from the following equation
\begin{equation}
\nu=1-\frac{\theta}{\pi}-\sum_{k=1}^{L}C_{2k}\frac{\sin(2k\theta)}{2k}.
\end{equation}
Note that $0\leq\nu\leq 1/2$ corresponds to  $-z\leq\mu\leq 0$ and  
$\pi\geq\theta\geq\pi/2$. Writing $\nu=1/2-\delta$ and $\theta=\pi/2+\eta$, 
Eqs.~(23) and (24) can then be rewritten as
\begin{equation}
E_{T}\left(\frac{1}{2}-\delta\right)
=-\frac{z}{\pi}\cos\eta+\frac{z}{2}\sum_{k=1}^{L}
(-1)^kC_{2k}\left[\frac{\cos[(2k-1)\eta]}{2k-1}
-\frac{\cos[(2k+1)\eta]}{2k+1}\right]
\end{equation}
and
\begin{equation}
\delta=\frac{\eta}{\pi}+\sum_{k=1}^{L}(-1)^kC_{2k}\frac{\sin(2k\eta)}{2k}.
\end{equation}

In the small $\delta$ limit (i.e., the electron filling $\nu$ is close to 
one half and $\eta$ is very small) $\sin(2k\eta)\simeq 2k\eta$; we then have
\begin{equation}
\eta\simeq\frac{\pi}{1+\pi\,{\cal C}(a,b,c)}\,\delta,
\end{equation}
where
$${\cal C}(a,b,c)=\sum_{k=1}^{L}(-1)^kC_{2k}(a,b,c).$$
Therefore, using the approximation 
$\,\cos[(2k\pm 1)\eta]\simeq 1-(2k\pm 1)^2\eta^2/2$, 
we obtain the total energy $E_{T}(1/2-\delta)$ close to 
(and below) half-filling
\begin{eqnarray}
E_{T}\left(\frac{1}{2}-\delta\right)&\simeq&-\frac{z}{\pi}\cos\left(\frac{\pi}
{1+\pi\,{\cal C}}\,\delta\right)+z\sum_{k=1}^{L}
\frac{(-1)^k}{(2k-1)(2k+1)}\,C_{2k}
+\frac{z}{2}\frac{\pi^2{\cal C}}{(1+\pi{\cal C})^2}\delta^2 \nonumber \\
&\simeq&E_T\left(\frac{1}{2}\right)+\frac{z}{\pi}\left[1-\cos\left(\frac{\pi}
{1+\pi\,{\cal C}}\,\delta\right)\right]
+\frac{z}{2}\frac{\pi^2{\cal C}}{(1+\pi{\cal C})^2}\delta^2.
\end{eqnarray}
Finally, by differentiating both $E_T$ in Eq.~(23) and $\nu$ in Eq.~(24) 
with respect to $\mu$, we can establish the following identity
\begin{equation}
\frac{dE_T}{d\mu}=\mu\frac{d\nu}{d\mu}.
\end{equation}
It is worthwhile to emphasize that the formalism described in this section 
is equally applicable to the 2D square lattice case.

\section{Magnetic moment and orbital susceptibility}
In this section we investigate two experimentally accessible observables, 
namely, the magnetic moment $M$ and the orbital susceptibility $\chi$ of 
this quantum system. It is well known\cite{kittel} that at absolute zero 
temperature $M$ and $\chi$ are given by the first- and second-order 
derivatives of the total energy 
\begin{equation}
M=-\frac{\partial E_T}{\partial B}
\end{equation}
and
\begin{equation}
\chi=-\frac{\partial^2 E_T}{\partial B^2}.
\end{equation}
With our analytical results for $E_T$ at hand, equations (30) and (31) 
provide a straightforward way to study these important quantities for any 
orientation of the externally applied field. For illustration purposes, 
below we will focus on the following flux orientations: $(0,0,\phi)$, 
$(0,\phi,\phi)$, $(\phi,\phi,\phi)$, and $(\phi,\phi,-2\phi)$. It will 
be seen that the magnetic response, in the presence of a strong periodic 
potential, is significantly different from the familiar Landau diamagnetism 
of a 2D electron gas, where $-\chi$ has its largest value at $B=0$ and 
its vicinity and decreases {\em monotonically} as $B$ is increased.

\subsection{Zero-field susceptibility $\chi(\mu)=\chi(\phi=0,\mu)$ versus 
Fermi energy $\mu$}
When we increase the electron filling factor $\nu$, the Fermi energy $\mu$  
increases. We therefore first study the zero-field susceptibility 
$\chi(\mu)=\chi(\phi=0,\mu)$ as a function of $\mu$ through
\begin{equation}
\chi(\mu)= \left.-\frac{\partial^2 E_T}{\partial B^2}\right|_{B=0}.
\end{equation}
>From Eq.~(23), we readily obtain
\begin{equation}
\chi(\mu)=\frac{z}{2}\sum_{k=2}^{L}
\left(\left.\frac{d^2C_{2k}(\phi)}
{d\phi^2}\right|_{\phi=0}\right)
\left[\frac{\sin[(2k-1)\theta]}{2k-1}
+\frac{\sin[(2k+1)\theta]}{2k+1}\right],
\end{equation}
where $d^2C_{2k}(\phi)/d\phi^2|_{\phi=0}$ can be easily computed for each of 
the flux orientations under consideration. 

In Fig.~3 we plot the negative of the susceptibility $-\chi$ as a function of 
the Fermi energy $\mu$ for various flux configurations. Large diamagnetic 
(i.e., negative) susceptibilities are observed at very low electron filling. 
For increasing $\mu$, the quantity $-\chi$ decreases and fluctuates around 
zero. For $\mu$ larger than a certain value ($\simeq -3.2$), paramagnetism 
prevails. Indeed, the orbital response is weakly paramagnetic at 
half-filling ($\mu=0$). These results are in qualitative agreement with 
the observations of Skudlarski and Vignale\cite{vignale} based on 
calculating the 
current-current correlation function. Furthermore, note that the absolute 
value of $\chi$ for $(0,\phi,\phi)$ is always twice that for $(0,0,\phi)$; 
$|\chi|$ for $(\phi,\phi,\phi)$ is always three times that for $(0,0,\phi)$; 
and $|\chi|$ for $(\phi,\phi,-2\phi)$ is twice that for $(\phi,\phi,\phi)$. 
These results clearly reflect the orbital origin of the susceptibility.

\subsection{Field-dependent $M(\phi)$ and $\chi(\phi)$ at $\nu=\frac{1}{2}$}

We now focus on the system at half-filling and examine the {\em continuous} 
dependence of the magnetic moment $M$ and susceptibility $\chi$ on the 
magnetic flux $\phi$. Employing Eq.~(20) for $E_T(\phi)$, we can directly 
obtain
\begin{equation}
-M(\phi)=\frac{dE_T(\phi)}{d\phi}
=z\,\sum_{k=0}^{L}\frac{(-1)^k}{(2k-1)\,(2k+1)}\,\frac{dC_{2k}(\phi)}{d\phi},
\end{equation}
and
\begin{equation}
-\chi(\phi)=\frac{d^2E_T(\phi)}{d\phi^2}
=z\,\sum_{k=0}^{L}\frac{(-1)^k}{(2k-1)\,(2k+1)}\,
\frac{d^2C_{2k}(\phi)}{d\phi^2}.
\end{equation}
The results for four different flux orientations are shown in Fig.~4. 
Let us look at $-M(\phi)=-M(\phi,\nu=1/2)$ first. Starting from $\phi=0$, 
$-M(\phi)$ decreases from zero and remains negative with some fluctuations. 
There are then some irregular ``sine-function-shaped" oscillations present. 
These low-field oscillations in $-M(\phi)$ are observed for all orientations 
shown in Fig.~4. For the $(0,0,\phi)$ flux orientation, $-M$ has a positive 
value for $\phi/2\pi\approx 0.3$--$0.5$, reaching zero at $\phi=\pi$. On the 
other hand, for $(0,\phi,\phi)$, $(\phi,\phi,\phi)$ and $(\phi,\phi,-2\phi)$, 
$-M$ is mostly negative for low fields ($\phi/2\pi\approx 0$--$0.1$), 
oscillates around zero for intermediate fields 
($\phi/2\pi\approx 0.1$--$0.3$), 
is negative for larger fields ($\phi/2\pi\approx 0.3$--$0.5$), and reaches 
zero at $\phi=\pi$. At relatively large flux values 
($\phi/2\pi\approx 0.4$--$0.5$), $-M$: decreases to zero (at $\phi=\pi$) 
from a positive value for $(0,0,\phi)$; and increases to zero 
(at $\phi=\pi$) from a negative value for $(0,\phi,\phi)$, 
$(\phi,\phi,\phi)$, and $(\phi,\phi,-2\phi)$. 

The field-dependent orbital magnetic susceptibility, 
$-\chi=-\chi(\phi,\nu=1/2)$, fluctuates somewhat evenly around zero in 
the low-field regime. However, $-\chi$ fluctuates less, and around a small 
negative value for the flux orientation $(0,0,\phi)$, where only one plane 
($xy$-plane) is penetrated by the flux. The fluctuations are more pronounced 
in the flux orientations: $(0,\phi,\phi)$, $(\phi,\phi,\phi)$, and 
$(\phi,\phi,-2\phi)$, where at least two perpendicular planes are affected 
by the field. As the flux is raised, $-\chi$ tends to fluctuate less and: 
(i) a crossover from diamagnetism to paramagnetism is observed in the 
configuration $(0,0,\phi)$; and, on the other hand, (ii) a crossover from 
paramagnetism to diamagnetism is observed in the orientations 
$(0,\phi,\phi)$, $(\phi,\phi,\phi)$, and $(\phi,\phi,-2\phi)$. 
In other words, at $\phi=\pi$, the magnetic response is weakly paramagnetic 
in the flux configuration $(0,0,\phi)$, while the other three orientations 
provide a strong diamagnetic response in the lattice. 

Figure~4 suggests that, for both $M$ and $\chi$, the following four 
quantities tend to grow larger with the increase of planes exposed to the 
field: the number of oscillations, the frequency of the oscillations as a 
function of the flux, the number of nodes, and the amplitude of the 
oscillations. For instance, $M(\phi,\phi,\phi)$ oscillates more rapidly and 
strongly than $M(0,\phi,\phi)$, which at the same time oscillates more 
rapidly and strongly than $M(0,0,\phi)$. A similar general trend is 
observed between $\chi(\phi,\phi,\phi)$, $\chi(0,\phi,\phi)$, and 
$\chi(0,0,\phi)$. These features can be understood intuitively: 
the more perpendicular planes 
are exposed to the flux, the more strongly $M$ and $\chi$ will be affected.

\section{Summary}
In conclusion, we present an investigation of quantum interference 
phenomena of tight-binding electrons on the 3D cubic lattice in a 
{\em continuously-tunable} magnetic field with {\em arbitrary orientation}. 
Previous work on this problem focused on a {\em discrete} set of (rational) 
magnetic field values. We study the total kinetic energy, and subsequently 
the magnetic moment and orbital susceptibility. The main results include: 
(1) an analytic study of electron quantum
interference effects resulting from sums over magnetic phase factors 
associated with 3D closed paths, 
(2) a very efficient computation of these ``lattice path-integrals" 
${\cal S}_{2l}$ in closed-form expressions, (3) explicit analytic 
expressions, in terms of the lattice path-integrals, for the 
Fermi-sea ground-state energy $E_T$ as a function of the fluxes for 
electron fillings $\nu$ at and near one half, (4) the 3D lattice 
path-integrals to very high order and the total energies in various 
flux orientations, (5) the investigation of the zero-field orbital 
susceptibility $\chi(\mu)$ as a function of the Fermi energy $\mu$, and (6) 
the magnetic moment $M(\phi)$ and susceptibility $\chi(\phi)$ as 
functions of the flux, both at half-filling.

We find that the absolute minimum of $E_T(\phi)$ at half-filling occurs 
at $\phi=\pi$ in all of the flux orientations under consideration, except 
for the configuration $(0,0,\phi)$ in 3D. It becomes evident that hopping 
in an additional direction drastically changes the properties found in 
strictly two-dimensional systems. It is also seen that the magnetic 
response, in the presence of a strong periodic potential, is significantly 
distinct from the familiar Landau diamagnetism---where in a 2D electron 
gas $-\chi$ takes the largest value close to $B=0$, and decreases 
monotonically with increasing $B$. For the zero-field susceptibility 
$\chi(\mu)$, diamagnetism dominates in spite of small fluctuations of 
$\chi(\mu)$ around zero for small electron filling $\nu$. On the average, 
with increasing $\mu$, the quantity $-\chi(\mu)$ {\em non-monotonically} 
decreases from a large positive (diamagnetic) value to a relatively 
small negative (paramagnetic) value. For $\mu\geq -3.2$, the orbital 
response becomes paramagnetic. Both the field-dependent 
$-M(\phi)$ and $-\chi(\phi)$ exhibit irregular oscillations according to the 
direction of the field. For the four flux orientations $(0,0,\phi)$, 
$(0,\phi,\phi)$, $(\phi,\phi,\phi)$, and $(\phi,\phi,-2\phi)$, 
the magnetic moment $M(\phi)$ is always zero at $\phi=0$ and $\pi$; 
and paramagnetism 
($\chi >0$) exists at $\phi=0$ for all these flux orientations. However, 
when $\phi/2\pi=1/2$, the magnetic response is paramagnetic for the 
flux configuration $(0,0,\phi)$, and diamagnetic for the other three 
orientations.

Here, we add a remark on the quantity $dE_T(\phi)/d\phi$. It is 
known that at zero temperature the persistent current\cite{landauer} 
in a metal ring threaded by a magnetic flux $\varphi$ is proportional to 
the sum over the contributions  of $\partial E_n/\partial\varphi$ from 
all occupied states, where $E_n$ is the eigenstate energy. We therefore 
can regard $dE_T(\phi)/d\phi$ as a generalized ``current" in this 
multiply-connected lattice system.

An approach commonly employed in recent years to study electrons 
in a magnetic field $\phi = p/q$ (e.g., Refs.~4-6) uses the 
Bloch theorem and maps the problem into a $ q \times q $ 
matrix problem---related to a 
$ q \times q $ cell with periodic boundary conditions (PBC) 
in the actual lattice. Thus, for $\phi = 1/2$, the electron energy 
levels are determined by considering a $2 \times 2$ cell with PBC 
in the lattice and diagonalizing a $2 \times 2$ matrix.  
This approach presents a problem for any irrational field, since 
$ q \rightarrow \infty $, and a periodic cell cannot be realized. 
We do not follow this approach, thus, we can explicitly consider a 
{\it continuum\/} of values of $\phi$, without treating irrational 
numbers on a special footing.  

Our lattice path-integrals are local 
quantities. By construction, they are valid for {\em any} value of the 
magnetic field.  
The PBC we impose [in Eq.~(9)] are only designed to make these lattice 
path-integrals homogeneous 
(i.e., translationally invariant) and have nothing to do with the 
imposition of a Bloch theorem.  Indeed, in contrast to most works 
studying electrons in a magnetic field, here we never invoke any k-space
or momentum-space: our sums over closed paths (lattice path-integrals) 
are all explicitly defined in real space.

The lattice path-integrals obtained here are 
``many-loop" generalizations of the 
standard ``one-loop" Aharonov-Bohm-type argument, where the electron
wave function picks up a phase factor $e^{i\Phi}$ each time it travels
around a closed loop enclosing a net flux $\Phi$. 
The evaluation of these lattice path-integrals 
enables us to analytically obtain the
total energies, magnetic moments, and orbital susceptibilities of the 
corresponding flux states. The spirit of
our approach follows Feynman's programme: to derive physical quantities
in terms of ``sums over paths". This method is considerably different
from the standard ones that have been employed so far (e.g., transforming 
the problem to momentum space and computing $E_T$ numerically). 
In particular, it allows the analytic calculation of physical quantities 
as explicit functions of a continuously-tunable flux, while other 
approaches need to separately consider different cases (e.g., matrices) 
for several discrete (rational) values 
of the magnetic field. 
The lattice path approach can also be used for a variety of other physical 
problems, including the derivation and analysis of the superconducting 
transition temperature in wire networks and Josephson-junction arrays 
(see, e.g., Refs.~10 and 11), and the analytical computation of the 
magnetoconductance for strongly localized electrons.\cite{lin1}

\acknowledgments
We thank A. Rojo for conversations and for bringing to our 
attention Ref.~7, and G. Vignale for useful comments on the manuscript. 
We thank O. Pla, supported by NATO grant CRG-931417, for his help. 
We acknowledge partial support from the University of Michigan 
Horace H. Rackham School of Graduate Studies, and the Offices of the 
Vice Presidents for Research and Academic Affairs.

\newpage

\begin{table}
\caption{Lattice path-integrals ${\cal S}_{2l}$ (with the order $2l=4, 6, 8
, 10, 12$) for the 3D cubic lattice in various flux configurations: 
$(0,0,\phi)$, $(0,\phi,\phi)$, $(\phi,\phi,\phi)$, $(\phi,\phi,-2\phi)$, 
and the asymmetric case $(\pi,\pi,\phi)$.}
\begin{tabular}{cl}
$2l$ & $\begin{array}{l} {\cal S}_{2l}\ \ {\rm in}\ \ (0,0,\phi) 
\end{array}$  \\ \hline
$4$ & $\begin{array}{l} 82+8\cos\phi \end{array}$ \\
$6$ &  $\begin{array}{l} 1452+384\cos\phi+24\cos2\phi \end{array}$ \\
$8$ &  $\begin{array}{l} 29218+13440\cos\phi+1960\cos2\phi+96\cos3\phi
+16\cos4\phi \end{array}$ \\
$10$ &  $\begin{array}{l} 638756+422800\cos\phi+96760\cos2\phi+11760\cos3\phi
+2284\cos4\phi+160\cos5\phi+40\cos6\phi \end{array}$  \\
\vspace{-0.1in} \\
$12$ &  $\begin{array}{l} 14865220+12737640\cos\phi+3909984\cos2\phi+
764216\cos3\phi+182328\cos4\phi  \\ 
+28392\cos5\phi+7680\cos6\phi+528\cos7\phi+144\cos8\phi+24\cos9\phi 
\end{array}$  \\
\hline \hline
$2l$ & $\begin{array}{l} {\cal S}_{2l}\ \ {\rm in}\ \ (0,\phi,\phi) 
\end{array}$  \\ \hline
$4$ & $\begin{array}{l} 74+16\cos\phi \end{array}$ \\
$6$ &  $\begin{array}{l} 1188+576\cos\phi+96\cos2\phi \end{array}$ \\
$8$ &  $\begin{array}{l} 22186+16736\cos\phi+4976\cos2\phi+736\cos3\phi
+96\cos4\phi \end{array}$ \\
$10$ &  $\begin{array}{l} 460236+461760\cos\phi+186360\cos2\phi+
49760\cos3\phi+11840\cos4\phi+2080\cos5\phi+520\cos6\phi \end{array}$  \\
\vspace{-0.1in} \\
$12$ & $\begin{array}{l} 10354220+12623616\cos\phi+6211104\cos2\phi
+2269840\cos3\phi+742800\cos4\phi  \\  
+209328\cos5\phi+67008\cos6\phi+14016\cos7\phi+3744\cos8\phi+480\cos9\phi 
\end{array}$  \\
\hline \hline
$2l$ & $\begin{array}{l} {\cal S}_{2l}\ \ {\rm in}\ \ (\phi,\phi,\phi) 
\end{array}$  \\ \hline
$4$ & $\begin{array}{l} 66+24\cos\phi \end{array}$ \\
$6$ &  $\begin{array}{l} 948+756\cos\phi+144\cos2\phi
+12\cos3\phi \end{array}$ \\
$8$ &  $\begin{array}{l} 16626+19392\cos\phi+6744\cos2\phi+1584\cos3\phi+
336\cos4\phi+48\cos5\phi \end{array}$ \\
$10$ &  $\begin{array}{l} 338616+483420\cos\phi+230340\cos2\phi+
82980\cos3\phi+27000\cos4\phi+7740\cos5\phi+1980\cos6\phi+420\cos7\phi+
60\cos8\phi \end{array}$  \\
\vspace{-0.1in} \\
$12$ &  $\begin{array}{l} 7672212+12250440\cos\phi+7095780\cos2\phi+
3290664\cos3\phi+1370952\cos4\phi+528120\cos5\phi+193524\cos6\phi  \\ 
+65736\cos7\phi+20952\cos8\phi+5952\cos9\phi
+1512\cos10\phi+288\cos11\phi+24\cos12\phi \end{array}$  \\
\hline \hline
$2l$ & $\begin{array}{l} {\cal S}_{2l}\ \ {\rm in}\ \ (\phi,\phi,-2\phi) 
\end{array}$  \\ \hline
$4$ & $\begin{array}{l} 66+16\cos\phi+8\cos2\phi \end{array}$ \\
$6$ & $\begin{array}{l} 912+528\cos\phi+336\cos2\phi
+48\cos3\phi+36\cos4\phi \end{array}$  \\
$8$ &  $\begin{array}{l} 14930+14016\cos\phi+10080\cos2\phi+3040\cos3\phi+
1992\cos4\phi+384\cos5\phi+240\cos6\phi+32\cos7\phi+16\cos8\phi \end{array}$ \\
\vspace{-0.1in} \\
$10$ &  $\begin{array}{l} 282796+356640\cos\phi+275560\cos2\phi+
122640\cos3\phi+79780\cos4\phi+28160\cos5\phi+17280\cos6\phi \\ 
+5280\cos7\phi+3040\cos8\phi+800\cos9\phi+
440\cos10\phi+80\cos11\phi+60\cos12\phi \end{array}$  \\
\vspace{-0.1in} \\
$12$ &  $\begin{array}{l} 6045696+9120912\cos\phi+7393704\cos2\phi
+4172032\cos3\phi+
2809680\cos4\phi+1342512\cos5\phi  \\ +841808\cos6\phi+367440\cos7\phi
+220836\cos8\phi+90384\cos9\phi
+52800\cos10\phi+19680\cos11\phi \\  +11856\cos12\phi+3696\cos13\phi
+2112\cos14\phi+576\cos15\phi
+360\cos16\phi+48\cos17\phi+24\cos18\phi
 \end{array}$ \\
\hline \hline
$2l$ & $\begin{array}{l} {\cal S}_{2l}\ \ {\rm in}\ \ (\pi,\pi,\phi) 
\end{array}$  \\ \hline
$4$ & $\begin{array}{l} 50+8\cos\phi \end{array}$ \\
$6$ &  $\begin{array}{l} 492+192\cos\phi+24\cos2\phi \end{array}$ \\
$8$ &  $\begin{array}{l} 5410+3456\cos\phi+808\cos2\phi+96\cos3\phi
+16\cos4\phi \end{array}$ \\
$10$ &  $\begin{array}{l} 64676+56720\cos\phi+18680\cos2\phi+4080\cos3\phi
+1000\cos4\phi+160\cos5\phi+40\cos6\phi \end{array}$  \\
\vspace{-0.1in} \\
$12$ &  $\begin{array}{l} 826820+900840\cos\phi+372384\cos2\phi+
113336\cos3\phi+35448\cos4\phi  \\ 
+9192\cos5\phi+2880\cos6\phi+528\cos7\phi+144\cos8\phi+24\cos9\phi 
\end{array}$ 
\end{tabular}
\label{table2}
\end{table}

\newpage
\noindent
{\bf FIGURE CAPTIONS}

\bigskip

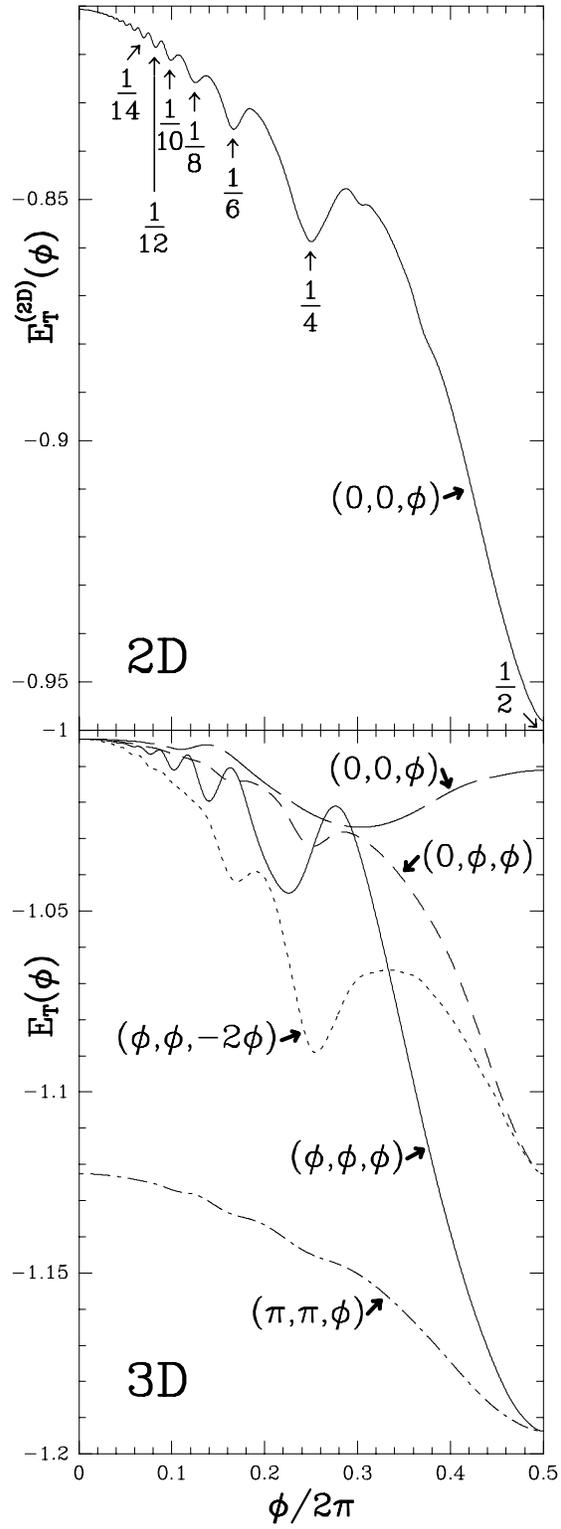
\begin{figure}
\caption{Lower frame: Total kinetic energy of electrons at half-filling 
$E_T(\phi)$ for the 3D cubic lattice for various flux orientations: 
$(0,0,\phi)$ (long dash), $(0,\phi,\phi)$ (short dash), $(\phi,\phi,\phi)$ 
(solid), $(\phi,\phi,-2\phi)$ (dot), and $(\pi,\pi,\phi)$ (dot-dash). 
Upper frame: $E^{(2D)}_T(\phi)$ for the 2D square lattice in a perpendicular 
field. The flux values where the minima occur are indicated. We obtain 
these results by using the lattice path-integrals up to the order $40$ 
for the 3D case and $76$ for the 2D lattice---which corresponds to summing 
contributions over $\sim 10^{29}$ electron closed paths in a 3D cubic 
lattice and $\sim 10^{44}$ 
electron closed paths in a 2D square lattice. Notice that $E_T(\phi)$ has 
the absolute minimum at $\phi/2\pi=1/2$ for all of these flux orientations, 
except for the configuration $(0,0,\phi)$.}
\label{fig1}
\end{figure} 

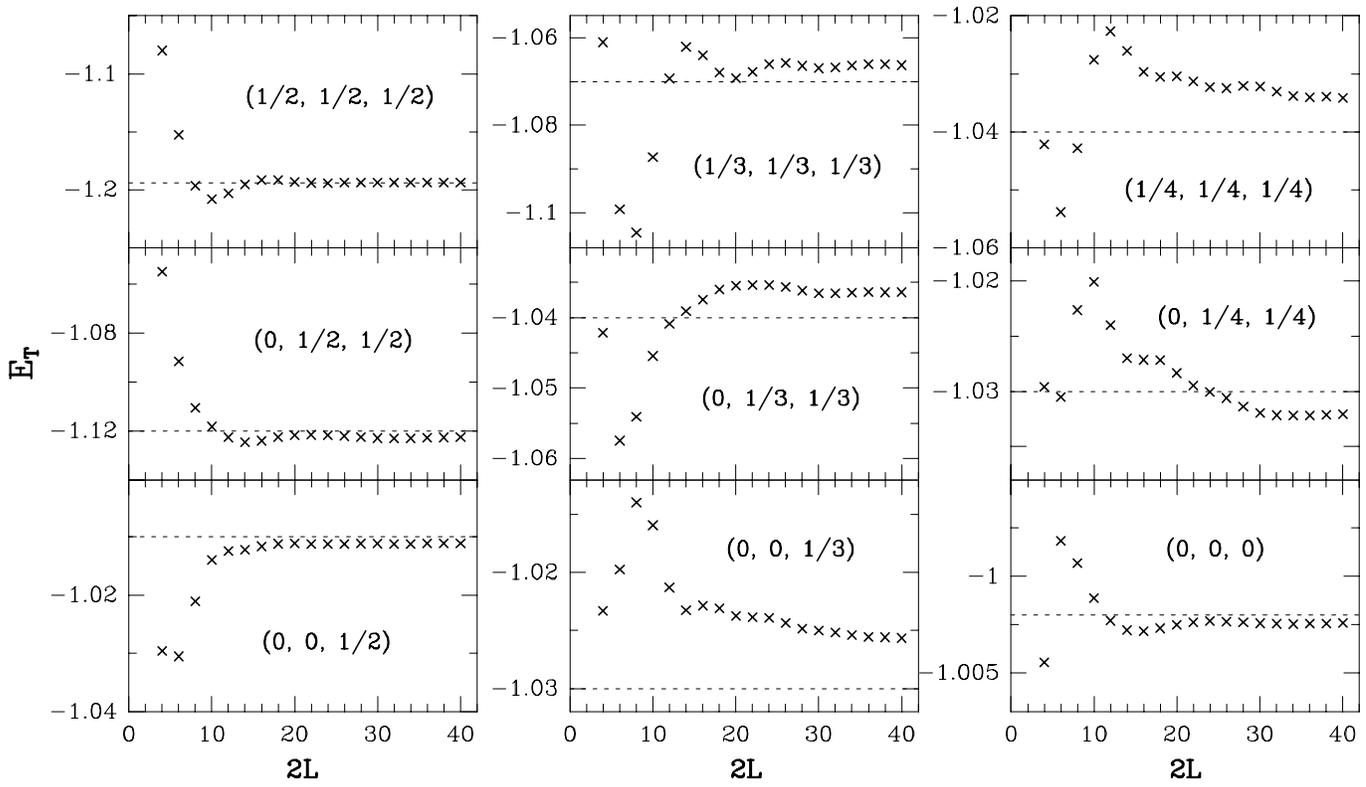
\begin{figure}
\caption{Total kinetic energy of electrons at half-filling $E_T$ for the 3D 
cubic lattice for various field orientations: $(1/2,1/2,1/2)$, 
$(1/3,1/3,1/3)$, $(1/4,1/4,1/4)$, $(0,1/2,1/2)$, $(0,1/3,1/3)$, 
$(0,1/4,1/4)$, $(0,0,1/2)$, $(0,0,1/3)$, and $(0,0,0)$. Here $(a,b,c)$ means 
that the three fluxes through the elementary plaquettes on the 
$yz$-, $zx$- and $xy$-planes are respectively $a/2\pi$, $b/2\pi$ 
and $c/2\pi$. The horizontal axis ($2L$) denotes the 
order of the highest-order lattice path-integral used in computing $E_T$. 
The dashed lines indicate the corresponding values from Ref.~5, 
where only the first three significant figures are available, except for 
the cases $(1/2,1/2,1/2)$ and $(0,0,0)$. The $E_T$ reach  steady values 
which are consistent, within $\sim 2\%$, with numerical ones for 
lattice path-integrals of order $2L$ equal to either $4$, $6$, or $8$, 
depending on the field orientation. Higher order 
lattice path-integrals provide negligible contributions to $E_T$. 
Notice that the vertical axes here cover very narrow range of values 
($\sim 10^{-2}$). For $(1/2,1/2,1/2)$, $(0,1/2,1/2)$ and $(0,0,1/2)$, 
both approaches give the same result within $\sim 1\%$ for $2L\geq 8$ and 
within $\sim 2\%$ for $2l\geq 8$, $8$, and $4$, respectively. For 
$(1/3,1/3,1/3)$, $(0,1/3,1/3)$ and $(0,0,1/3)$, both approaches provide 
the same result within $\sim 1\%$ for $2L\geq 12$, $10$, and $12$, and 
within $\sim 2\%$ for $2l\geq 8$, $4$, and $4$, respectively. For 
$(1/4,1/4,1/4)$, $(0,1/4,1/4)$ and $(0,0,0)$, both approaches yield 
the same result within $\sim 1\%$ for $2L\geq 16$, $4$, $4$ and within 
$\sim 2\%$ for $2l\geq 4$, $4$, and $4$, respectively.}
\label{fig2}
\end{figure} 

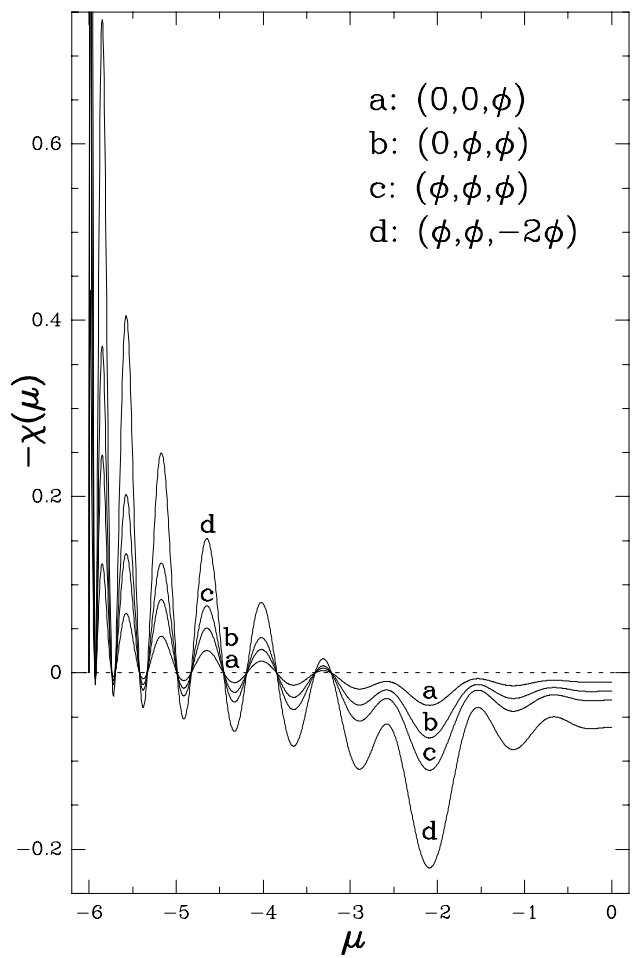
\begin{figure}
\caption{The negative of the zero-field susceptibility $-\chi(\mu)$ as a 
function of the Fermi energy $\mu$ for various flux configurations: 
$(0,0,\phi)$, $(0,\phi,\phi)$, $(\phi,\phi,\phi)$, and 
$(\phi,\phi,-2\phi)$ corresponding to curves $a$, $b$, $c$, and $d$, 
respectively. Curve $a$ is always the closest to the dotted reference line 
$\chi=0$, while curve $d$ is always the farthest from it. For all these 
orientations, the susceptibility exhibits a very non-monotonic behavior. 
For small electron filling $\nu$ (i.e., small $\mu$), 
diamagnetism ($\chi <0$) dominates in spite of small fluctuations of 
$-\chi$ around zero. With increasing $\mu$, 
on the average, $-\chi$ decreases from a large positive value to a 
negative one. For $\mu\geq -3.2$ the orbital response is paramagnetic.}
\label{fig3}
\end{figure}

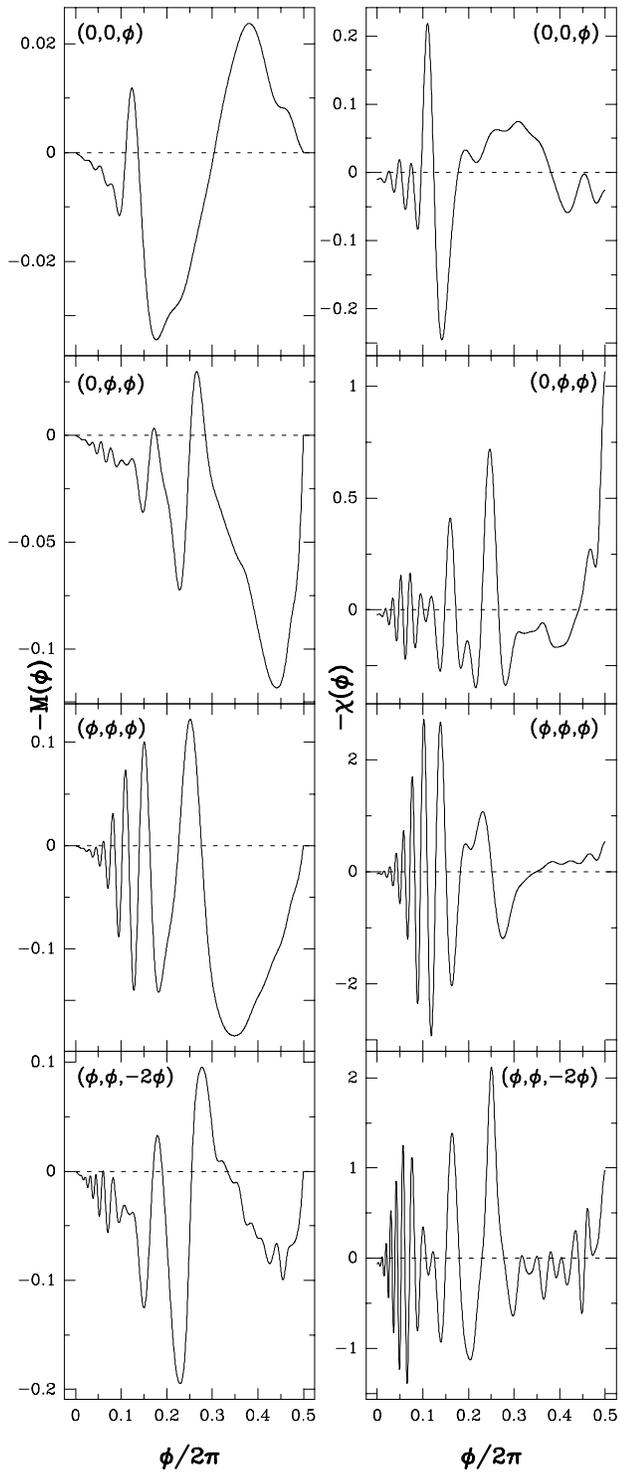
\begin{figure}
\caption{The negatives of the magnetic moment $-M(\phi)$ (left column) and 
the orbital susceptibility $-\chi(\phi)$ (right column) at half-filling as 
a function of the flux $\phi$ for various flux orientations as indicated. 
>From top to bottom they are: $(0,0,\phi)$, $(0,\phi,\phi)$, 
$(\phi,\phi,\phi)$, and $(\phi,\phi,-2\phi)$. Both $-M$ and $-\chi$ exhibit 
non-periodic oscillations. Notice the change of the amplitude and the 
frequency of the oscillations as a function of the flux for different 
orientations of the field. At $\phi=0$ and $\pi$, the magnetic moments are 
always zero. Moreover, paramagnetism exists at $\phi=0$ for all these 
flux orientations. However, when $\phi/2\pi=1/2$, the magnetic response 
is paramagnetic for flux configuration $(0,0,\phi)$, and diamagnetic 
for the other three orientations.}
\label{fig4}
\end{figure}


\begin{references}
\bibitem[*]{lyl}Present address: Department of Physics, West Virginia 
University, Morgantown, West Virginia 26506-6315. 
\bibitem[\dag]{fnori}Electronic address: nori@umich.edu 
\bibitem{hlrw}See, e.g., I. Affleck and J. B. Marston, 
Phys. Rev. B {\bf 37}, 3774 (1988); Y. Hasegawa {\em et al.}, 
Phys. Rev. Lett. {\bf 63}, 1519 (1989); P. Lederer, D. Poilblanc, 
and T. M. Rice, {\em ibid.} {\bf 63}, 1519 (1989); D. Poilblanc, 
Y. Hasegawa, and T. M. Rice, Phys. Rev. B {\bf 41}, 1949 (1990); 
F. Nori, E. Abrahams, and G. T. Zimanyi, {\em ibid.} {\bf 41}, 7277 (1990); 
M. Kohmoto and Y. Hatsugai, {\em ibid.} {\bf 41}, 9527 (1990); 
F. Nori, B. Dou\c cot, and R. Rammal, {\em ibid.} {\bf 44}, 7637 (1991); 
and references therein.
\bibitem{nori}F. Nori and Y.-L. Lin, Phys. Rev. B {\bf 49}, 4131 (1994).
\bibitem{lieb}E. H. Lieb, Phys. Rev. Lett. {\bf 73}, 2158 (1994).
\bibitem{vignale}P. Skudlarski and G. Vignale, Phys. Rev. B {\bf 43}, 5764 
(1991).
\bibitem{hasegawa}Y. Hasegawa, J. Phys. Soc. Jpn. {\bf 59}, 4384 (1990).
\bibitem{zee}Z. Kunszt and A. Zee, Phys. Rev. B {\bf 44}, 6842 (1991).
\bibitem{rice}W. F. Brinkman and T. M. Rice, Phys. Rev. B {\bf 2}, 1324 (1970).
\bibitem{kittel}C. Kittel, {\em Introduction to Solid State Physics}, 
6th ed. (John Wiley \& Sons, New York, 1986); and N. W. Ashcroft 
and N. D. Mermin, {\em Solid State Physics} (Saunders College, 
Philadelphia, 1976).
\bibitem{landauer}R. Landauer and M. Buttiker, Phys. Rev. Lett. 
{\bf 54}, 2049 (1983).
\bibitem{niu}Q. Niu and F. Nori, Phys. Rev. B {\bf 39}, 2134 (1989).
\bibitem{lin}Y.-L. Lin and F. Nori, Phys. Rev. B {\bf 50}, 15953 (1994).
\bibitem{lin1}Y.-L. Lin and F. Nori, Phys. Rev. Lett. {\bf 76} 4580 (1996); 
Phys. Rev. B {\bf 53}, 15543 (1996).
\end{references}
\end{document}